\documentclass[aps,prl,showpacs,12pt]{revtex4}
 
\usepackage{graphicx}

\newcommand{\beq}{\begin{equation}}
\newcommand{\eeq}{\end{equation}}

\begin{document}

\title{The role of the stress trap in polariton quasiequilibrium condensation in GaAs microcavities}

\author{R. Balili}
\author{B. Nelsen}
\author{D.W. Snoke}
\affiliation{Department of Physics and Astronomy, University of Pittsburgh\\
3941 O'Hara St., Pittsburgh, PA 15260, USA}

\author{L. Pfeiffer}
\author{K. West}
\affiliation{Bell Labs, Lucent Technologies\\
700 Mountain Ave., Murray Hill, NJ 07974, USA}


\date{}

\begin{abstract}
Recent experiments have shown several effects indicative of Bose-Einstein condensation in polaritons in GaAs-based microcavity structures when a harmonic potential trap for the two-dimensional polaritons is created by applied stress. These effects include both real-space and momentum-space narrowing, first-order coherence, and onset of linear polarization above a particle density threshold. Similar effects have been seen in systems without traps, raising the question of how important the role of the trap is in these experiments. In this paper we present results for both trapped conditions and  resonant, non-trapped conditions in the same sample. We find that the results are qualitatively different, with two distinct types of transitions. At low density in the trap, the polaritons remain in the strong-coupling regime while going through the threshold for onset of coherence; at higher density, there is a different threshold behavior which occurs with weak coupling and can be identified with lasing; this transition occurs both with and without a trap. 

\end{abstract}

\maketitle

\newpage
Several recent papers \cite{deveaud,science,yama,vortex,love,baum,superfluid} have shown dramatic effects of spontaneous coherence of polaritons in microcavities. The polariton in these systems consists of a bosonic superposition of a quantum well exciton and a cavity photon; the polarizability of the exciton leads to strong coupling of the exciton and photon states when their energies are resonant \cite{kav-review}.  The  coupling to the cavity photon mode gives the polaritons extremely light mass ($\sim 10^{-4}  ~m_0$, where $m_0$ is the vacuum electron mass), which implies that quantum degenerate effects can occur at moderate densities and high temperatures; indeed, one system has been explored with polaritons at room temperature \cite{baum-RT}.  The excitonic part of the polaritons gives them a short-range polariton-polariton interaction which is orders of magnitude stronger than the photon-photon interaction in a typical optical medium. These characteristics mean that the polariton gas is well modeled as a weakly-interacting, two-dimensional boson gas with extremely light mass.  

Because the mirrors of the microcavity are leaky, however, the polaritons have a short lifetime, ($\sim 5$ ps). In a typical experiment, a  steady state or quasi-steady state population of polaritons is maintained by incoherent optical or electrical pumping. Questions therefore remain as to how well the phase transitions of these quasiparticles under different conditions can be described as Bose-Einstein condensation (BEC).  Of course, the absolute time scale does not matter; what matters is the thermalization time compared to the lifetime.  At high density, polaritons can collide with each other on subpicosecond time scales, allowing the particles to approach equilibrium within their lifetime. If the density is too high, however, phase-space filling of the valence and conduction bands can set in, removing the strong coupling of the photonic and electron states, which in turn means that one can no longer think of the system as elastically scattering bosonic particles, and instead must view it as lasing of photons amidst an incoherent electron-hole plasma.  
\begin{figure}
\includegraphics[width=0.55\textwidth]{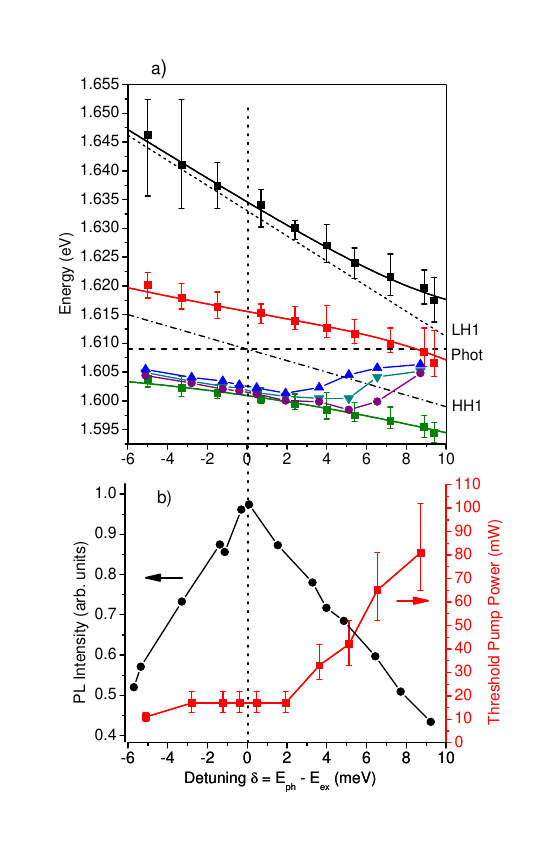}
\caption{a) Squares: energy of the reflection minima of the cavity as a function of detuning,  when stress is applied to vary the exciton energy while leaving the cavity photon energy unchanged (increasing stress = increasing $\delta$). The data are fit to the model of coupled states discussed in the text, using the exciton energies (HH1, LH1) and cavity photon energy (Phot) shown. Circles: the peak photon emission energy as a function of detuning when a laser excites the sample, with power at the threshold for spectral narrowing. Inverted triangles: the photon emission energy when the laser excitation power is increased by a factor of 1.7 beyond the threshold. Upright triangles: the photon emission energy when the laser excition power is increased by a factor of 2.5 beyond the threshold.  b) Circles, left axis: Photoluminescence intensity of the lower polariton line as a function of detuning, for laser excitation density well below threshold (9 mW, with spot size 85 microns). Squares, right axis: The laser power needed to reach the threshold for coherent behavior (corresponding to the power used for the circles in (a).) Laser spot size was 30 $\mu$m; laser photon energy was 716 nm, at the absorption at the top edge of the microcavity stop band.}
\end{figure}

In our recent experiments \cite{science}, we trapped polaritons in an in-plane harmonic potential created by applied stress.  A harmonic potential has the advantage that it makes true Bose condensation possible in two dimensions \cite{nozieres,berman} and reduces the total number of particles needed for coherent effects, by changing the density of states of the particles.  In the limit of zero spring constant, the condensate fraction vanishes \cite{berman,baym} but the superfluid fraction is nonzero (the Berezinsky-Kosterlitz-Thouless transition). As the spring constant of the trap is increased, a condensate fraction and the superfluid fraction both increase at a given temperature \cite{berman}; in other words, the total particle density needed for spontaneous coherence is reduced.   

When inhomogeneous applied stress is used to shift the excitonic states, a harmonic potential is produced in real space which gives a drift force that confines the polaritons \cite{berman}.  It has been argued for a similar system \cite{deveaud} that a random potential arising from disorder effectively also makes an in-plane trap which can confine the particles and allow true BEC. The random potential severely inhibits long-range motion of the particles, however. In our experiments with GaAs structures, the disorder is very low, and the polaritons can move tens of microns  and approach spatial equilibrium in a macroscopic trap.

The experiments with trapped polaritons \cite{science} showed several effects associated with BEC, namely 1) spatial condensation in the center of the trap, even when the laser generated the polaritons far from that point, 2) momentum-space narrowing into a bimodal distribution, 3) sudden occurrence of linear polarization, and 4)  first-order coherence.  Although these all indicate that the phase transition is analogous to BEC, an objection can be raised. The polariton densities at which these effects occurred is not so much less than the density at which a lasing transition can be seen without a trap in nearly identical GaAs-based structures \cite{bloch}. Does the presence of the trap make such a difference, that the character of the phase transition is in the strong coupling regime, when the particle density is only about a factor of three or four lower than the density at which a transition to standard lasing occurs in the weak coupling regime when there is no trap? The answer, surprisingly, is yes. 

A key way to learn about the nature of the transition is to see how the energy of the states varies as the detuning of the cavity is varied. If the system is in the weak coupling regime, then the light-emitting state should be essentially the same as the cavity photon, and therefore the emission energy should follow the cavity photon energy as the detuning varies.  This is what was observed in Ref. \cite{bloch} and what we observe when no stress is applied (as in Ref. \cite{bloch}, the cavity photon energy above the lasing threshold is red shifted relative to the bare photon energy, presumably due to renormalization of the dielectric constant.) When stress is applied to create the in-plane trap, however, we observe that the emitted photon energy at the threshold for coherence follows the lower polariton state as it shifts downward with stress due to the shift of the exciton state. This shows that the polaritons remain in the strong coupling regime even above the threshold.

The squares in Figure 1(a) show the energy positions of the polaritonic states in a microcavity, as the detuning between the exciton states and the cavity photon energy is changed by varying the applied stress, using the method discussed in Ref.~\cite{apl}.  The sample is the same as that used for Ref.~\cite{science}, and substantially the same as those used in Refs.~\cite{yama-pnas} and \cite{bloch}.  The positions of the lines are well fit with a simple three-state coupling model, namely the eigenvalues of the matrix
\begin{equation}
H = \left(  \begin{array}{ccc}
E_{\rm HH1} & 0 &  \Omega_1\\
0 & E_{\rm LH1} & \Omega_ 2\\
\Omega_1 & \Omega_2 & E_{\rm phot} 
\end{array}\right),
\end{equation}
with $\Omega_1 = 7.5$ meV, $\Omega_2 = 6.0$ meV, $E_{\rm phot} = 1.609$ eV, and  $E_{\rm HH1}$ and $E_{\rm LH1}$ shifting with stress as shown.  The shear term of the deformation potential Hamiltonian acts to decrease the splitting of the heavy and light hole states in our stress configuration, unlike the case of a homogeneous uniaxial stress.   The line positions are consistent with the reported masses \cite{lb} of the light and heavy holes, i.e., the heavy hole exciton energy in the quantum wells, $E_{\rm HH1} \propto (1/m_e + 1/m_h) = (1/0.067m_0 + 4/0.33m_0)$, and light hole exciton energy, $E_{LH1} \propto (1/m_e + 1/m_l) = (1/0.067m_0 + 1/0.094m_0)$, with a well width of 61~\AA.  Both exciton states couple to the cavity mode when they are near resonance.  At the resonance of the HH1 exciton state and the cavity photon, i.e. at zero detuning, the photoluminescence intensity (PL) has a maximum, as seen in Figure 1(b). As discussed by Stanley et al. \cite{stanley}, there is a fine structure of the PL intensity near the zero detuning point, which we also observe, but which is not seen here because of the spacing of the stress points in this case. The full width at half maximum of the PL intensity resonance around $\delta = 0$ is about 10 meV, the same resonance width as seen when there is no stress and the photon energy is tuned, instead of the exciton energy, by varying the location of the laser spot on the sample. 
 
 \begin{figure}
\includegraphics[width=0.6\textwidth]{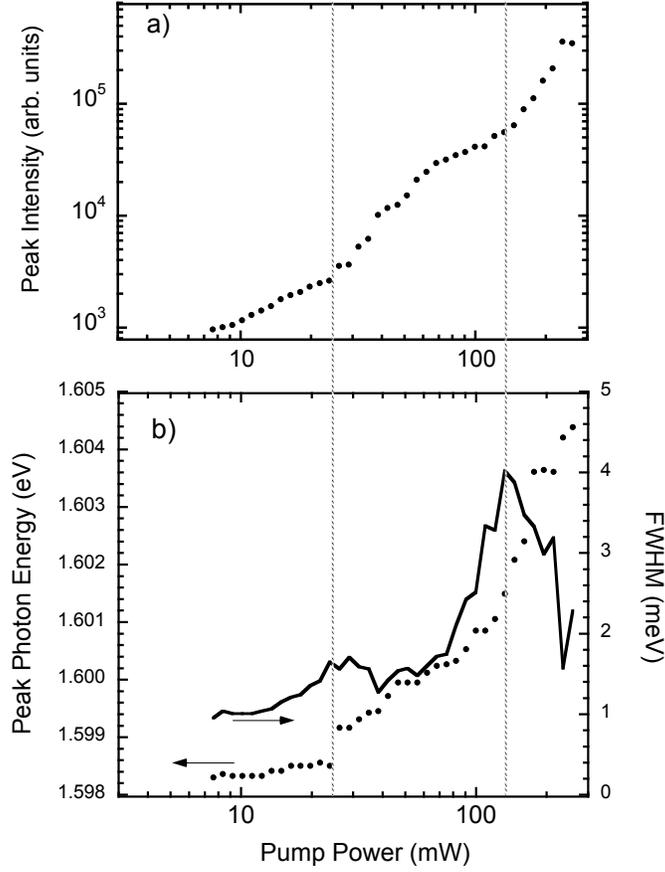}
\caption{a) Peak intensity of the emission from the lower polariton as a function of pump power when the system is at zero detuning, when the polaritons are generated in a stress trap under the same conditions as those for Fig.~1.  b) Dots, left axis: peak photon energy of the emission for the same conditions as (a). Solid line, right axis: the full width at half maximum of the emission spectrum under the same conditions. The acceptance angle for the PL detection was $0 \pm 3^{\circ}$. A different region of the sample was used, so that the lower polariton energy at zero detuning in this case is around 1.5984 eV, as compared to 1.600 eV in Fig.~1.}
\end{figure}

\begin{figure}
\includegraphics[width=0.6\textwidth]{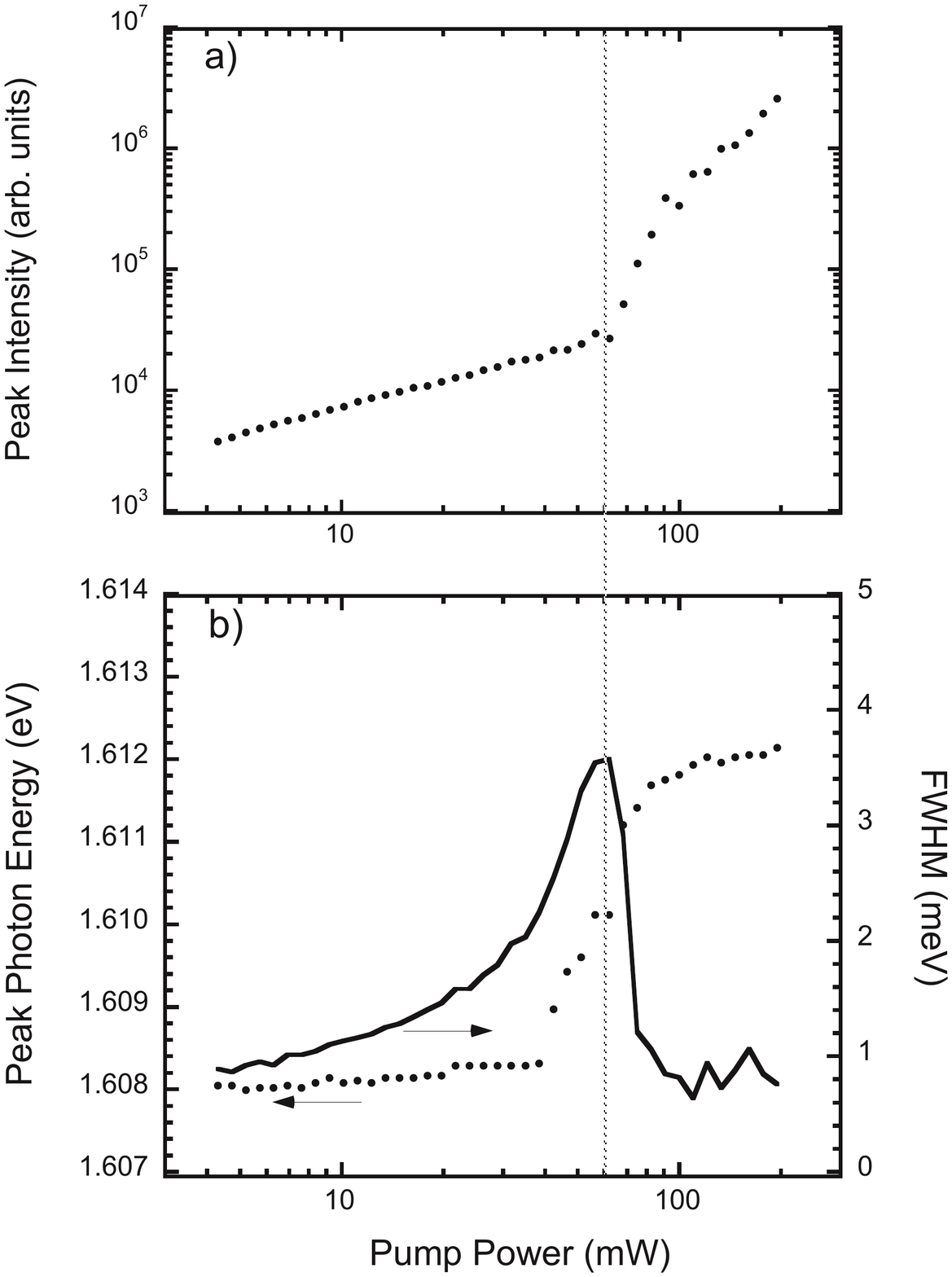}
\caption{a) Peak intensity of the emission from the lower polariton as a function of pump power when the system is at zero detuning, when there is no stress trap--- a location is chosen such that the exciton and cavity photon states are in resonance.  b) Dots, left axis: peak photon energy of the emission for the same conditions as (a). Solid line, right axis: the full width at half maximum of the emission spectrum under the same conditions. The acceptance angle for the PL detection was the same as for Fig.~2.}
\end{figure}

Figure 1(a) also shows the energy of the photon emission when a laser pumps the sample under conditions  similar to those in Ref.~\cite{science}, i.e., the laser photon energy is tuned to the first absorption maximum above the microcavity stopband, and the laser is circularly polarized.  The circles correspond to the photon energy when the excitation density is exactly at the threshold for coherent effects, which include line narrowing and a nonlinear increase of the emission intensity. The inverted and upright triangles correspond to laser powers which are higher than the threshold power by ratios of 1.7 and 2.5, respectively.  These data show that even well above the threshold for the coherent effects, the lower polariton energy follows the exciton energy, not the photon energy, until the system is quite detuned. When it reaches detunings larger than $\delta =  4$ meV or so, the emission photon energy jumps up to near the bare cavity photon energy. At this same point, as shown in Fig.~1(b), the power needed to cause coherent behavior increases rapidly.  At this point we conclude that the system is in weak coupling.

It thus appears that there are two distinct transitions occurring in the same sample. The lower-power threshold can be identified with Bose condensation of polaritons in the strong coupling limit, and occurs only when the trap exists, while the higher threshold can be identified with standard lasing in the weak coupling regime, and occurs in the unstressed sample as well as in the stressed sample when it is detuned away from resonance.

This identification is supported by examining what happens at zero detuning when the pump power is increased. In this case we expect two transitions at the same place. First, we expect to see the lower, strong-coupling condensation transition, and then as power is increased, we expect to see the weak coupling lasing transition kick in when the excitation density is comparable to that of the weak-coupling transition in the unstressed case. This is indeed what we see.  Figure 2 shows the peak intensity, peak energy, and full width at half maximum (FWHM) of the emission line as a function of pump power for the
stressed case at zero detuning, while Figure 3 shows the same data for the case when there is no stress trap, at zero detuning.  The two cases are quite different. In the unstressed case shown in Fig.~3, the line narrowing and nonlinear emission do not occur until the emission line has shifted almost 4 meV, putting the system close to the weak coupling regime. In the case with the stress trap, line narrowing occurs at much lower power, when the line shift is only about 0.5 meV. As seen in Fig.~2(b), the line width and shift remain around this plateau, and the intensity gain saturates, until the density increases by a factor of 4, at which point the line broadens again, and the blue shift of the line jumps up several meV. A second threshold of line narrowing occurs, along with a second range of nonlinear increase of the peak intensity, which we attribute to lasing in the weak coupling regime. This second transition was not seen in the data of Ref.~\cite{science} because the maximum pump intensity was less in those experiments. In both cases, the nonlinear gain region, which also corresponds to the region of narrowest line width, occurs over a range of density about a factor of three above the critical density. Above that, the light emission quickly begins to broaden and shift to higher energy. 

The fact that the narrowing at the lower, polaritonic threshold is only about 25\% can be attributed to the fact that our pump laser is a multimode laser with significant fluctuations on nanosecond time scales. This causes a fluctuating shift of the line position which is recorded by our time-integrating detection system as a broadened line. As shown by Love et al. \cite{love}, when an intensity-stabilized laser is used, very narrow line widths ($\sim 0.05$ meV) and long coherence times ($\sim 150$ ps) are recorded for this type of polaritonic transition. 
\begin{figure}
\includegraphics[width=0.45\textwidth]{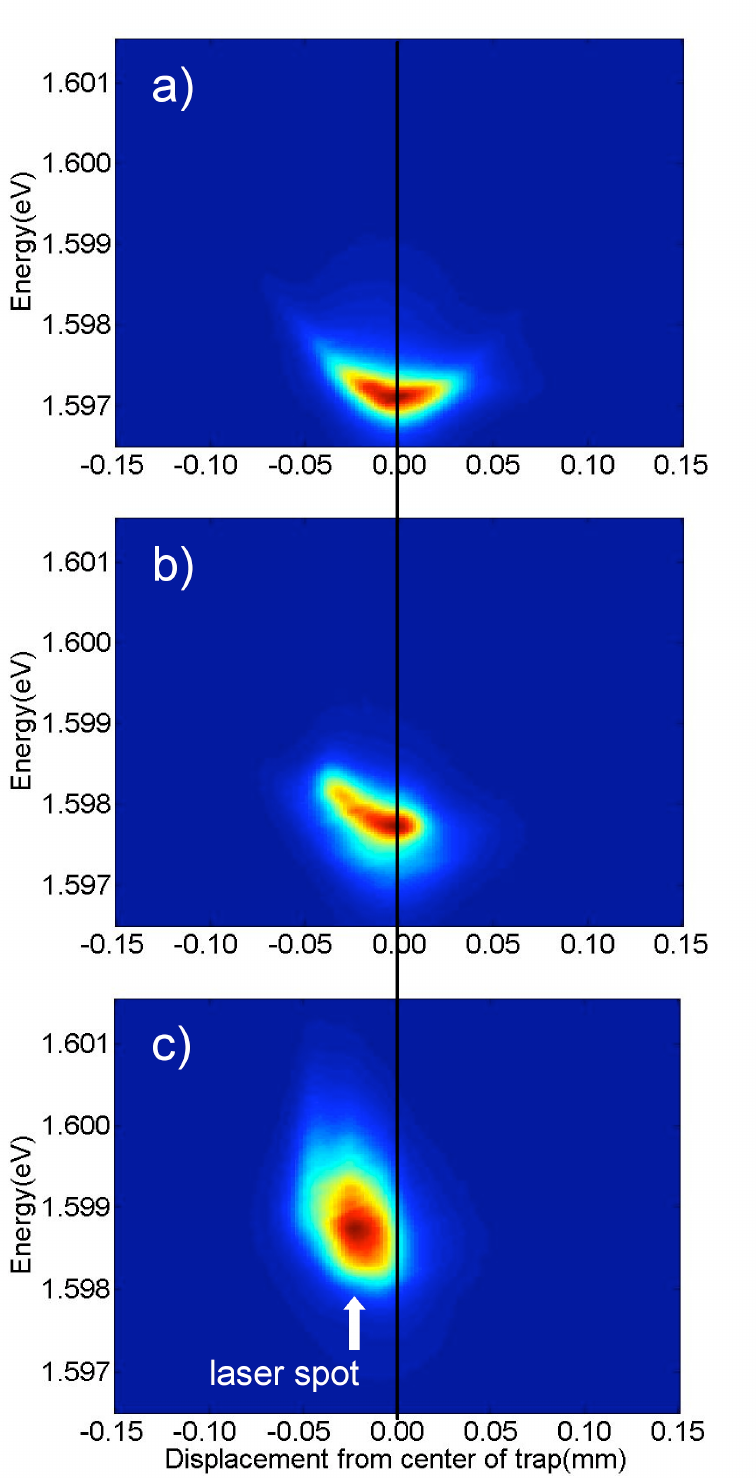}
\caption{a) Spatially-resolved spectrum of the light emission from polaritons in the trap, when the polaritons are at low density, well below the threshold for coherent behavior, in the case when the detuning $\delta = + 1$ meV. The laser was quasi-cw with 2.4\% duty cycle and average power 0.1 mW, corresponding to 4.2 mW while the laser was on.  b) Spatially-resolved spectrum at the first critical threshold for coherent behavior and spectral narrowing (average laser power 0.6 mW, corresponding to 25 mW while the laser was on.)  c) Well above the critical threshold (average power 1.6 mW). The excitation wavelength and polarization for all three images were nearly the same as for the data of Figs. 1 and 2, but the laser was focused on a spot not at the center of the trap. A different region of the sample was used, so that the lower polariton energy at zero detuning in this case is 1.597 eV.}
\end{figure}

A final confirmation of the different nature of the two transitions seen in the spatially-resolved spectra shown in Fig.~4. For this experiment, the laser was not focused at the center of the trap, but off to the side, as indicated by the arrow. As seen in Fig.~4(a), at low power the polaritons collect in the center of the trap. Above the first critical threshold, there is a spatial narrowing as well as a spectral narrowing of the light emitted from the center of the trap, as seen in Fig.~4(b). This behavior is consistent with a condensation of polaritons in the center of the trap. As the excitation density is increased further, approaching the second threshold, the center of the light emission moves to the center of the laser excitation spot, as shown in Fig.~4(c). At this higher excitation density, the diffusion of the polaritons is much reduced. This is the same density at which the exciton luminescence spectrum begins to shift to the blue and to broaden.  Thus the picture of weakly interaction BEC is breaking down at this density. Eventually, if the density is increased enough, the spectrum narrows again and standard lasing occurs at the same spot. The trap plays no role, as the photons in the cavity are amplified at the point of generation of the carriers.  

We have found the diffusion length of the polaritons to be highly sensitive to the average power of the pump laser, an effect which we attribute to lattice heating. In the case of Fig.~ 4(a), a low duty-cycle pulsed excitation was used, to keep the average laser power low, allowing the lattice to be cold enough that the polaritons can flow to the center of the trap at low density. The data of Figs. 1 and 2 were recorded with continuous excitation at the center of the trap. The instantaneous powers needed to reach the threshold for coherent behavior are comparable in the two cases, however.

In conclusion, the lasing transition, in which the carriers are in a plasma state and the photons are weakly coupled to the carriers, and the polariton condensate transition, in which the photons and excitons are strongly coupled to make bosonic polaritons, are clearly distinguishable, even though both lead to emission of coherent light.  The polaritonic coherence clearly occurs when the excitonic component of the polaritons is important, as seen by the shift of the lower polariton emission at the threshold with stress to follow the bare exciton state. The Rabi splitting between the upper and lower polariton remains large, indicating that phase space filling is not significant. By contrast, the lasing transition occurs when the splitting between the upper and lower polaritons has closed up, so that the emission is near the bare cavity photon energy.  The trap plays an essential role in making the polariton condensate transition possible.  If there is no trap, only the lasing transition can be seen in these samples.  If there is a trap, both transitions can occur. When the excitation is displaced from the center of the trap, lasing occurs at the excitation spot, while the polariton condensation seeks to occur at the center of the trap. The two transitions can occur at carrier densities which are less than a factor of ten different.  This should not be a surprise, because the Mott transition can have a sudden onset \cite{exion,barjoseph}.  Once a Mott transition occurs, only lasing of a plasma in the weak coupling regime can occur. 

The stress trap used in these experiments appears to reduce the critical threshold for polaritonic coherence just enough to move it from above the Mott transition density to below it.  As discussed above, the trap has a key role in making BEC possible in a 2D system; in a two-dimensional flat potential, fluctuations will deplete the condensate. Another effect of the trap which may play a role in these experiments is simply that the trap gives the excitons higher density by collecting them in the center of the trap. While the the exciton density can also be increased by simply turning up the pump laser power, increased laser power also leads to increased lattice heating. The trap helps to produce a colder, denser gas.   Numerical models \cite{hartwell} indicate that the applied stress may also play an indirect role in creating free carriers via ionization of impurities due to the piezoelectric effect. These free carriers may help to thermalize the polaritons to the lattice temperature.

{\bf Acknowledgments}. This work has been supported by the National Science Foundation under grant DMR-0706331.  We thank J. Bloch and C. Weisbuch for helpful conversations, and V. Hartwell for early contributions to these experiments.

\end{document}